\documentclass[useAMS,usenatbib,twocolumn]{mn2e}
\usepackage{amsfonts}
\usepackage{amsmath}
\usepackage[dvips]{graphicx}
\usepackage{epsfig}
\usepackage{enumitem}   
\usepackage{wrapfig}
\usepackage{amssymb}
\usepackage{longtable}
\usepackage{accents}
\usepackage{dcolumn}
\usepackage{multirow}
\usepackage{bm}
\usepackage[bottom]{footmisc}
\usepackage{url}
\usepackage[caption=false]{subfig}
\usepackage{enumitem}   
\SetLabelAlign{CenterWithParen}{\hfil{#1}\hfil}  

\begin{document}

\title[Radio structure of TXS 0506+056]{VLBI radio structure and radio brightening of the high-energy neutrino emitting blazar TXS 0506+056}
\author[E. Kun, P. L. Biermann, L. \'{A}.
Gergely]{E. Kun$^{1}$\thanks{%
E-mail: kun@titan.physx.u-szeged.hu}, P. L. Biermann$^{2,3,4,5}$, L. \'{A}. Gergely$^{1}$ \\
$^{1}$ Institute of Physics, University of Szeged, D\'om t\'er 9, H-6720 Szeged, Hungary\\
$^{2}$ Max Planck Institute for Radioastronomy, Auf dem H\"{u}gel 69, D-53121 Bonn, Germany\\
$^{3}$ Department of Physics, Karlsruhe Institute for Technology, P.O. Box 3640, D-76021, Karlsruhe, Germany\\
$^{4}$ Department of Physics \& Astronomy, University of Alabama, AL 35487-0324, Tuscaloosa, USA\\
$^{5}$ Department of Physics \& Astronomy, University of Bonn, Regina-Pacis-Weg 3, 53113, Bonn, Germany}
\date{Accepted . Received ; in original form }
\maketitle

\begin{abstract}
We report on the radio brightening of the blazar TXS 0506+056 (at $z=0.3365$), supporting its identification as source of the high-energy (HE) neutrino IC-170922A by the IceCube Neutrino Observatory. MOJAVE/VLBA data indicate its radio brightness abruptly increasing since January 2016. When decomposing the total radio flux density curve (January 2008 - July 2018) provided by the Owens Valley Radio Observatory into eight Gaussian flares, the peak time of the largest flare overlaps with the HE neutrino detection, while the total flux density exhibits a threefold increase since January 2016. We reveal the radio structure of TXS 0506+056 by analysing VLBI data from the MOJAVE/VLBA Survey. The jet-components maintain quasi-stationary core separations. The structure of the ridge line is indicative of a jet curve at the region $0.5\div2$ mas ($2.5\div9.9$ pc projected) from the VLBI core. The brightness temperature of the core and the pc-scale radio morphology support a helical jet structure at small inclination angle ($<8\fdg2$). The jet pointing towards the Earth is key property facilitating multimessenger observations (HE neutrinos, $\gamma$- and radio flares). The radio brightening preceding the detection of a HE neutrino is similar to the one reported for the blazar PKS 0723--008 and IceCube event ID5.
\end{abstract}

\pagerange{\pageref{firstpage}--\pageref{lastpage}}

\label{firstpage}


\section{Introduction}

\begin{table*}
\begin{center}
\caption{Gaussian decomposition of the flux density curve of TXS0506+056 observed with the OVRO 40m telescope. The amplitude ($A$), time of peak flux density ($T_p$), and FWHM ($\theta$) of the flares (F1..f8) are listed in separate columns.}
\setlength\tabcolsep{4pt}
\begin{tabular}{ccccccccc}
\hline
\hline
 & F1 & F2 & F3 & F4 & F5 & F6 & F7 & f8\\
\hline
$A$ (Jy) & $ 0.59 \pm 0.03 $ & $ 0.16 \pm 0.03 $ & $ 0.08 \pm 0.05 $ & $ 0.39 \pm 0.02 $ & $ 0.37 \pm 0.04 $ & $ 0.328 \pm 0.017 $ & $ 0.65 \pm 0.01 $ & $ 0.46 \pm 0.03 $ \\
$T_{\rm p}$ (yr) & $ 2008.72 \pm 0.13 $ & $ 2009.91 \pm 0.02 $ & $ 2010.96 \pm 0.03 $ & $ 2011.86 \pm 0.19 $ & $ 2014.04 \pm 0.02 $ & $ 2015.35 \pm 0.02 $ & $ 2017.56 \pm 0.05 $ & $ 2018.52 \pm 0.03 $ \\ 
$\theta$ (yr) &$ 1.14 \pm 0.19 $ & $ 0.30 \pm 0.05 $ & $ 0.06 \pm 0.03 $ & $ 1.28 \pm 0.27 $ & $ 0.60 \pm 0.05 $ & $ 0.37 \pm 0.02 $ & $ 1.24 \pm 0.07 $ & $ 0.27 \pm 0.02 $ \\ 
\hline
\hline
\label{table:flarefit}
\end{tabular}
\end{center}
\end{table*}

\begin{table}
\begin{center}
\caption{Summary of the 15 GHz image parameters. (1) epoch of the VLBA observation, (2) VLBA experiment code, (3)--(4) FWHM major and minor axis of the restoring beam, respectively, (5) position angle of the major axis of the restoring beam measured from North to East, (6) rms noise of the image after the model-fit procedure. The full table is available in electronic format on-line.}
\centering
\setlength\tabcolsep{4pt}
\begin{tabular}{cccccccc}
\hline
\hline
Epoch	&	VLBA Code & $\mathrm{B}_{\mathrm{maj}}$ & $\mathrm{B}_{\mathrm{min}}$& $\mathrm{B}_{\mathrm{PA}}$ & rms\\
 &  & (mas) & (mas) & ($^{\circ}$) & (mJy bm$^{-1}$)\\
(1) & (2) &  (3) &  (4) &  (5) &  (6)\\
\hline
2009-01-07 & BL149 & 1.396 & 0.635 & -3.9 &  0.153\\
2009-06-03 & BL149 & 1.341 & 0.584 & -6.9 &  0.182\\
2010-07-12 & BL149CL & 1.385 & 0.509 & -8.5 & 0.175\\
2010-11-13 & BL149CW & 1.369 & 0.576 & -9.1 &  0.148\\
2011-02-27 & BL149DC & 1.175 & 0.536 & -0.8 &  0.210\\
\hline
\hline
\label{table_impars}
\end{tabular}
\end{center}
\end{table}

With the discovery of the cosmic high-energy (HE) neutrinos, extensive studies targeted their sources. Active galactic nuclei (AGN), driven by supermassive black holes (SMBHs), are prominent candidates, able to produce the dominant isotropic neutrino background between $10^4\div10^{10}$ GeV \citep{Stecker1991}.

Blazars, a subclass of AGN Urry1995, are characterized by a relativistic jet at small inclination angle, appearing much brighter than they would be intrinsically due to relativistic boosting and beaming of their electromagnetic radiation. They are highly variable sources, many of them radiating from radio frequencies to the $\gamma$-ray regime. Very long baseline interferometry (VLBI) observations exposed outward moving bright spots of the surface brightness distribution in a number of jets, that propagate with apparently superluminal speeds \citep[e.g.][]{Vermeulen1994,Lister2016}. Such superluminal motion arises from projection effects. 
 
A major $\gamma$-outburst of the blazar PKS B1424--418 has been reported by \citet{Kadler2016}, found to occur in temporal and positional coincidence with a PeV-energy shower-type neutrino event (ID35) with average median angular error $15\fdg9$, observed by the IceCube Neutrino Observatory.

Cross-correlating available radio catalogues with the arrival direction of the track-type neutrinos detected by the IceCube \citep{IC2014,IC2015,Schoenen2015}, we identified the blazar PKS 0723-008 as the source of the HE neutrino event ID5 \citep{Kun2017}. The positional identification was much stronger as for track-type events the median angular error of the parent neutrino is $\lesssim 1\fdg2$.
Along with this blazar, three other candidates were found \cite[the corresponding HE neutrino events are given in brackets;][]{Kun2017}: the BL Lac object PKS B1206--202 ($z=0.404$, ID8), the quasar PKS B2300--254 ($z$, ID18), the quasar PKS B2224+006 (4C+00.81; $z=2.25$, ID44). 

Recently the \citet[][]{ICTXS2018a} reported on the multimessenger observations of TXS 0506+056 \citep[$z=0.336$,][]{Paiano2018}. The HE neutrino event IceCube-170922A detected at 22/9/2018 was in temporal and positional coincidence with a prominent $\gamma$-ray flare. The deposited energy was $\sim290$ TeV in a track-type event. \citet{Padovani2018} investigated multiband archival monitoring data of TXS0506+056 from radio to $\gamma$-rays \cite[for mid-infrared archival data see][]{Gabanyi2018} and suggested the source underwent a hadronic flare with HE neutrino emission. Knowing the position of TXS 0506+056 as prior, the \citet[][]{ICTXS2018b} also identified an excess of HE neutrinos from the direction of this AGN between September 2014 and March 2015.

Our aim was to investigate the radio characteristics of the jet of TXS 0506+056 by employing archival VLBI data and considering the full Owens Valley Radio Observatory (OVRO) flux density curve, that starts at 8/1/2008 and ends at 24/7/2018. For VLBI analysis we explored the radio interferometric data produced by the "Monitoring Of Jets in Active galactic nuclei with VLBA Experiments" (MOJAVE) survey \citep{Lister2009}.

\begin{figure}
  \includegraphics[scale=0.6]{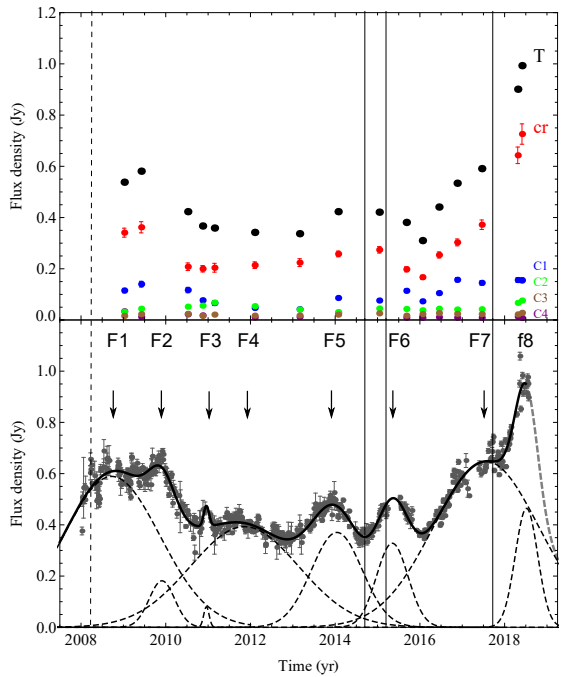}
 \caption{Upper: total flux density (T) of TXS 0506+056 from VLBI observations (MOJAVE/VLBA), as well as flux density of its jet components C1, C2, C3, C4, the VLBI core (cr) are plotted against time. Lower: flux density curve of TXS 0506+056 observed with the OVRO $40$ m single dish radio telescope (gray dots with errorbars), superimposed with its Gaussian decomposition into eight flares. The peak of the flares is indicated with their labels, where the number after 'F' is increasing with time of peak flux density. The beginning and ending of the HE neutrino excess identified between 2014 and 2015, as well as the detection of IC-170922A are indicated by vertical lines. The starting date of monitoring data at IceCube is indicated by a dashed vertical line \citep[the first date in sample IC40][]{ICTXS2018b}.}
 \label{figure:flareplot}
 \end{figure}

\begin{table}
\begin{center}
\caption{Circular Gaussian model-fit results for TXS~0506+056. (1) epoch of observation, (2) flux density $S$, (3)-(4) position of the component center respect to the core ($x$ and $y$), (5) FWHM major axis ($d$), (6) jet-component identification. The full table is available in electronic format on-line with precision to three decimal places.}
\setlength\tabcolsep{3pt}
\begin{tabular}{lccccc}
\hline
\hline
Epoch & $S$ & $x$ (mas) & $y$ (mas) & $d$  & CO\\ 
(yr) & (Jy) & (mas) & (deg) & (mas)  & \\ 
(1) & (2) & (3) & (4) & (5)  & (6)\\ 
\hline
2009.02 & $0.34 \pm 0.0$ & $0.00 \pm 0.02$ & $0.00 \pm 0.03$ & $0.08 \pm 0.01$ & cr \\
 & $0.12 \pm 0.01$ & $-0.13 \pm 0.04$ & $-0.37 \pm 0.04$ & $0.23 \pm 0.02$ & C1 \\
 & $0.03 \pm 0.01$ & $-0.52 \pm 0.05$ & $-1.66 \pm 0.08$ & $0.76 \pm 0.05$ & C2 \\
 & $0.03 \pm 0.00$ & $0.05 \pm 0.06$ & $-2.24 \pm 0.11$ & $1.07 \pm 0.02$ & C3 \\
 & $0.02 \pm 0.00$ & $0.94 \pm 0.12$ & $-3.92 \pm 0.29$ & $2.36 \pm 0.14$ & C4 \\
\hline
\hline
\label{partable}
\end{tabular}
\end{center}
\end{table} 

\section{Radio flux density variations of TXS 0506+056}

The MOJAVE programme \citep{Lister2009,Lister2013,Lister2016} provides archival calibrated data of TXS 0506+056 at 15 GHz, that were obtained for 16 epochs between $2009.016$ and $2018.412$. The measurements were performed with the Very Long Baseline Array (VLBA) interferometer. It consists of $10$ radio telescopes each having a diameter of $25$ m. The array employs the VLBI technique, reaching a resolution better than $1$ milliarcsecs (mas) at $15$ GHz. 

We plotted the total flux density of TXS 0506+056 from MOJAVE/VLBA data in the upper panel of Fig. \ref{figure:flareplot}. VLBI data indicate the radio brightness of the source is increasing since January
2016. We show the total flux density curve of TXS 0506+056 observed with the OVRO $40$-m single dish radio telescope at $15$ GHz in the lower panel of Fig. \ref{figure:flareplot}. The figure suggests a close correspondence between the OVRO and the VLBA flux density curves, indicating that the source is highly core dominated at $15$ GHz, with minimal kpc-scale radio emission. Minimizing the $\chi^2$, we decomposed it into $8$ Gaussian flares following the procedure discussed in \citet{Pyatunina2006} and \citet{Pyatunina2006} \cite[see other examples of the Gaussian decomposition of flux density curves e.g. in][]{Britzen2010,Kudryavtseva2011}. The amplitude ($A$), time of the peak-flux density ($T_{\rm p}$) and full width at half maximum (FWHM, $\theta$) duration of the individual the individual flares are summarized in Table \ref{table:flarefit}. 

The two largest flares ($A > 0.5$~Jy) peaked at $2008.72\pm0.13$ (F1) and $2017.56\pm0.05$ (F7), respectively. Moderately large flares peaked (0.3~Jy~$< A <$~0.5~Jy) at $2011.86\pm0.19$ (F4), $2014.04\pm0.02$ (F5), $2015.35\pm0.02$ (F6), respectively. The two smallest flares peaked at $2009.91\pm0.02$ (F2), $2010.96\pm0.03$ (F3) respectively. Though we fitted flare f8 with a Gaussian, this flare might be still in its brightening phase therefore its indicated peak time should be considered with caution.

The IceCube Collaboration reported on the identification of an excess of cosmic neutrinos in the position of TXS 0506+056 between September 2014 and March 2015 \citep[][]{ICTXS2018b}, overlapping with radio flare F6. The radio emission of TXS 0506+056 started to increase after the HE neutrino-excess, in the beginning of 2016, and produced a 3-fold increase in flux density until the most recent OVRO observations. The peak time of flare F7, which has the largest amplitude of the fully sampled flares overlaps with the detection time of the HE neutrino event IceCube-170922A (Fig. \ref{figure:flareplot}).

\section{VLBI radio structure of the jet}

  \begin{figure}\centering
 \includegraphics[scale=0.35,angle=270]{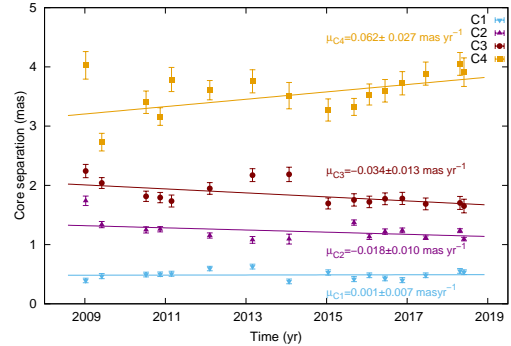}
 \caption{Core separation of components C1, C2, C3 and C4 as a function of time. The solid lines and the corresponding slopes represent their linear proper motion ($\mu$). }
 \label{figure:fitvel}
 \end{figure}
   \begin{figure}\centering
 \includegraphics[scale=0.35,angle=270]{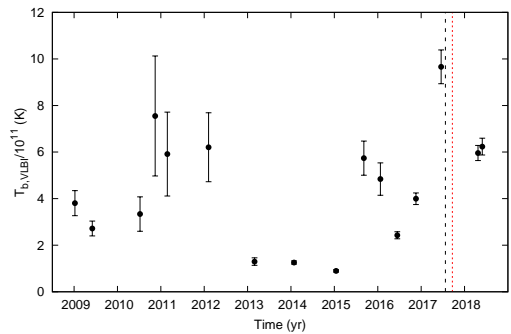}
 \caption{VLBI brightness temperature of the core as function of time. The time of peak flux density is marked by black dashed vertical line, and the detection time of HE neutrino IC-170922A is marked by red dotted vertical line.}
 \label{figure:tvlbi}
 \end{figure}
 
   \begin{figure*}
 \includegraphics[scale=0.51]{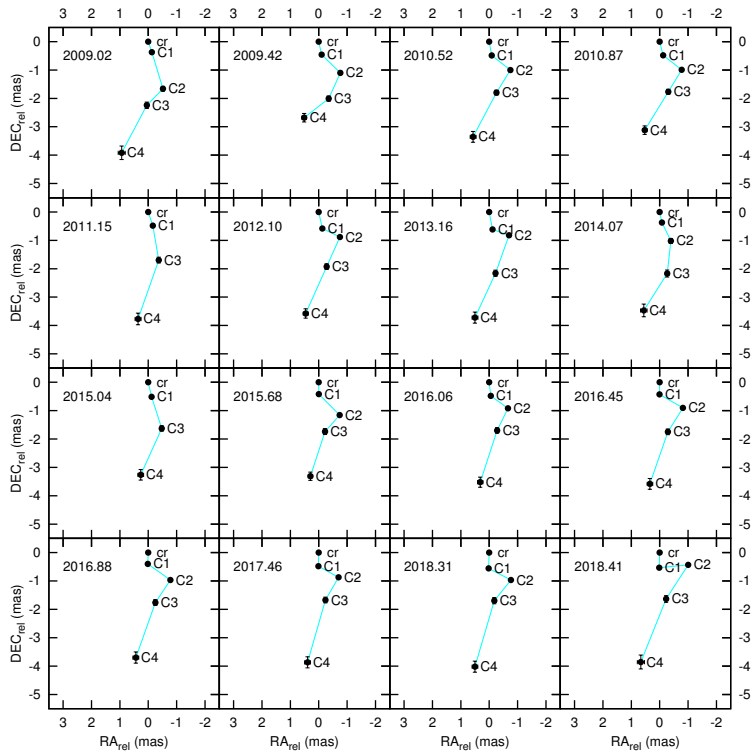}
 \caption{RA-DEC coordinates of components C$1$, C$2$, C$3$, C$4$ relative to the core position (0,0), which is marked by 'cr'. The observing epochs are marked in upper right corner of the maps. In four epochs a faint additional component can be seen (2009.42, 2010.52, 2010.87, 2015.68).}
 \label{figure:all_xy}
 \end{figure*} 
 
\subsection{Analysis of archival MOJAVE/VLBA observations}

We used standard \textsc{Difmap} tasks \citep{Shepherd1994} to model-fit the VLBI data, such that circular Gaussian components were employed to build up the surface brightness distribution of the jet. Table \ref{table_impars} shows the the summary of the image parameters. We identified four components, C$1$, C$2$, C$3$, C$4$, in a region extending up to $4$ mas in projection from the core \citep[$\approx 20$ pc, using the cosmological parameters from][]{Planck2015}. Component C1 (C4) has the smallest (largest) average distance to the VLBI core. The total flux density, sky position and width of the jet components are available in electronic format on-line, along with their observing epochs. The error of these parameters were estimated as in \citet{Kun2014}. 

\subsection{Structural evolution and orientation of the VLBI jet} 
 
We plotted the core separation of the jet components against their observing epochs with the best-fit slopes of their linear proper motion in Fig. \ref{figure:fitvel}. Components C1 and C2 reside quasi-stationary across the observations. C3 showed some inward-directed, while C4 showed some outward-directed motion. The confidence level for the fits are $4.39\sigma$ (C1), $>5\sigma$ (C2), $2.94\sigma$ (C3), $4.02\sigma$ (C4). They might have oscillatory motion, similar to that of the jet of S5 1803+784 \citep[][]{Britzen2010,Kun2018}. This is reflected by the bad fit of the linear proper motion--model to the data. We note that in case of TXS 0506+056 more observation are needed to see how the component behave in a longer time-span.

We plotted the right ascension ($x$) and declination ($y$) of the brightness peak of the jet components relative to the position of the VLBI core in Fig. \ref{figure:all_xy}. The average and standard deviation of the FWHM major and minor axis of the restoring beam is ($1.269\pm0.082$) mas and ($0.544\pm0.035$) mas, respectively. A jet curve appears in the region roughly between $0.5$--$2$ mas ($2.5$--$9.9$ pc) from the core, indicating helical jet ridge line, such that the jet components reside quasi-stationary.  

Due to Doppler boosting, the apparent brightness temperature $T_\mathrm{b}$ can exceed than the limiting intrinsic brightness temperature $T_\mathrm{int}$. The Doppler factor $\delta$ connects them as $T_\mathrm{b}=\delta T_\mathrm{int}$. The brightness temperature can be given in case of VLBI components as \citep[e.g.][]{Condon1982}:
\begin{flalign}
T_\mathrm{b,VLBI}=1.22\cdot 10^{12}\times(1+z) \frac{S_{\nu}}{d^2 \nu^2} \mbox{~}\mathrm{(K)},
\end{flalign}
where $S_{\nu}$ is the flux density (in Jy), $d$ is the FWHM diameter of the component (in mas), $\nu$ is the observing frequency (in GHz) and $z$ is the red-shift of the source. We estimated the apparent core-brightness temperature from the total flux density and diameter of the respective Gaussian component at each of the $16$ epochs of MOJAVE data. We plotted the VLBI brightness temperature of the core in Fig. \ref{figure:tvlbi}. The core VLBI brightness temperature was low during the HE neutrino excess observed between September 2014 and March 2015 and it was the highest nearly at the time of the peak flux density of the largest flare (F7), when HE neutrino IC170922A was also detected.
Equating the intrinsic brightness temperature with the equipartition brightness temperature $T_\mathrm{eq} \approx 5\times10^{10}$~K \citep{Readhead1994}, we obtained $\bar{\delta}=8.98\pm1.93$ for the core.
 
The lower limit on Doppler factor of the core $\bar{\delta}=8.98\pm1.93$ limits the Lorentz factor as $\gamma_{\rm min}=0.5 \times (\delta+\delta^{-1})$, resulting in characteristic jet speed $\beta$ larger than $\approx0.96c$, and limits the inclination angle as $\iota_{\rm max}=\arcsin(\delta^{-1})$, resulting in $\iota$ smaller than $\approx 8\fdg15$. The smaller (larger) value of the intrinsic temperature would yield larger(smaller) limit on the jet speed $\beta$ and smaller (larger) limit on the inclination angle $\iota$.

The small inclination angle of the jet and the lack of obvious globally outward-directed motion of all of the jet components suggest they can be identified as "lantern-regions", highly Doppler-boosted parts of the jet. Two of the present authors and their collaborators explained the apparent brightness variations leading to the appearance of lantern regions as caused by variable inclination along a helical ridge line \cite{Kun2018}. This was based on a model of the intrinsic ridge line \citep{Steffen1995} relying on conservation laws for kinetic energy, momentum and jet opening angle. The quasi-stationary feature of the components of S5 1803+784 has been explained as follows: the plasma reaches the respective Doppler-boosted regions, then moves out, fading below the sensitivity of the interferometer.  
 
\section{Discussion and summary}
\label{discussion}


In our earlier work \citep{Kun2017}, we have searched for the source of the track-type HE neutrinos detected by the Antarctic IceCube Neutrino Observatory \citep{IC2014,IC2015,Schoenen2015,ICTXS2018a}. For track-type events induced by a muon neutrino, the median angular error is $\sim1\fdg2$. We remarked that binary SMBH evolution qualitatively explains the observed HE neutrino emission through the following sequence of events: (1) the dominant black hole spin-flips \citep[presumably million years before the actual merger,][]{Gergely2009}, (2) a new jet-channel plows through the surrounding material capturing seed particles, also feeding from the accretion disk, (3) the dominant jet accelerates ultrahigh-energy cosmic ray particles (UHECRs), and the most energetic proton collisions generate pions that decay further, producing neutrinos, (4) the SMBH merger occurs with accompanying emission of low-frequency gravitational waves.

The dominant jet reoriented after the spin-flip is responsible for the observable consequences of the merger: (i) increased radiation in all EM frequencies due to the high Lorentz factor of the freshly made jet, (ii) flat spectrum up to THz frequencies due to the energetic synchrotron-radiating electrons \cite[see the Fermi LAT spectrum in][]{ICTXS2018a}, (iii) increasing radio flux density due to enhanced synchrotron radiation (this Letter), (iv) emission of HE particles \cite[neutrinos, UHECRs, $\gamma$-rays;][]{ICTXS2018a}. The identification of the neutrino sources is enhanced by the strong Doppler boosting of the jet, therefore the flat-spectrum emission is a key selection criterion to find source candidates. 

Considering the object of study of the present paper, the blazar TXS~0506+056, the post-merger SMBH binary scenario provides a consistent physical picture explaining the already occurred multimessenger observations (radio and $\gamma$ flaring, HE neutrinos). In particular, the coincident radio brightening (OVRO) and HE neutrino detection is similar to the one reported for the blazar PKS 0723-008 and IceCube event ID5 \citep{Kun2017}.

Decomposing the total flux density of TXS~0506+056 observed with the OVRO 40-m single dish radio telescope at 15 GHz into 8 Gaussian flares, the peak flux density of the largest flare (F7) coincided with the detection of the HE neutrino IC-170922A. Calculating the VLBI brightness temperature of the core based on its flux density and width, it was found the highest nearly at the time of the peak flux density of the flare F7. 

We also analysed its VLBI jet, finding that its radio structure, beyond the radio brightening, exhibits a prominent jet curve at 15 GHz, also that its components reside at quasi-stationary core separations. Such a property was already identified in the jet of the BL Lac object S5 1803+784 \citep{Kun2018} and interpreted as “lantern regions” in a helical jet, indicative of small inclination angles with respect to the line-of-sight. The average Doppler factor of the core limits the inclination angle as being smaller than $8\fdg2$. The radio jet pointing towards the Earth facilitates the emission of HE neutrinos in coincidence with the observed $\gamma$-ray flare \citep{ICTXS2018a} and the recent substantial increase in the radio emission of the blazar TXS~0506+056.
\section*{Acknowledgements}
We would like to thank the referee for valuable comments. EK thanks to K. {\'E}. Gab{\'a}nyi and S. Frey for the discussions on the topic. This work was supported by the Hungarian National Research Development and Innovation
Office (NKFIH) in the form of the grant 123996 and based upon work from the COST action CA15117 (CANTATA), supported by COST (European Cooperation in Science and Technology). This research has made use of data from the OVRO 40-m monitoring programme (Richards, J. L. et al. 2011, ApJS, 194, 29) which is supported in part by NASA grants NNX08AW31G, NNX11A043G, and NNX14AQ89G and NSF grants AST-0808050 and AST-1109911. This research has made use of data from the MOJAVE database that is maintained by the MOJAVE team (Lister et al., 2009, AJ, 137, 3718). The National Radio Astronomy Observatory is a facility of the National Science Foundation operated under cooperative agreement by Associated Universities, Inc.

\end{document}